\documentclass{aamas2016_extendedabstract}
\usepackage{algorithm}
\usepackage{algpseudocode}
\usepackage{paralist}
\usepackage{subcaption}
\usepackage{mathtools}
\DeclarePairedDelimiter\ceil{\lceil}{\rceil}

\begin{document}


\title{Multi-Agent Continuous Transportation with Online Balanced Partitioning}



%
%
%
%

%

\numberofauthors{1}

\author{
%
\alignauthor
Chao Wang\textsuperscript{\dag}, Somchaya Liemhetcharat\textsuperscript{\S},  Kian Hsiang Low\textsuperscript{\dag}\\
\affaddr{Department of Computer Science, National University of Singapore, Singapore\textsuperscript{\dag}}\\
\affaddr{Institute for Infocomm Reserach, Agency for Science, Technology and Research, Singapore\textsuperscript{\S}}\\
\email{\{wangchao, lowkh\}@comp.nus.edu.sg\textsuperscript{\dag}, liemhet-s@i2r.a-star.edu.sg\textsuperscript{\S}}
}

\maketitle

\begin{abstract}
We introduce the concept of \textit{continuous transportation} task to the context of multi-agent systems. A continuous
transportation task is one in which a multi-agent team visits a number of fixed locations, picks up objects, and delivers
them to a transportation hub. The goal is to maximize the rate of transportation while the objects are replenished over time
. In this extended abstract, we present a hybrid of centralized and distributed approaches that minimize
communications in the multi-agent team. We contribute a novel online partitioning-transportation algorithm with information gathering in the multi-agent team.
\end{abstract}



%
%

%
%
\printccsdesc






\section{Introduction}
We are interested in multi-agent coordination and continuous transportation, where agents visit locations in the
environment to transport objects (passengers or items) and deliver them to a transportation hub. The objects
replenish over time, and we consider Poisson model of object replenishment. Poisson model is suitable for scenarios with
independently-occurring objects, such as passengers appear at the transit stops. Thus, the continuous
transportation task is general and applicable to many real-life scenarios, e.g., first/last mile problem, package or mail
collection and delivery.

The most similar work has been done is in multi-robot foraging and delivery problem \cite{liemhetcharat2015continuous,liemhetcharat2015multi}.
A multi-robot team forages resources from environment to a home location or delivers items to locations on request while the resources replenish over time.
The goal of continuous foraging is to
maximize the rate of resource foraging. Thus, continuous foraging has similarities to continuous transportation, and thus we
compare our algorithms with that of \cite{liemhetcharat2015continuous, liemhetcharat2015multi}.

In addition, continuous area sweeping (e.g., \cite{ahmadi2006multi}) problem has similarities to continuous transportation, with the
main difference being that the agents have a carrying capacity and must periodically return to the transportation hub.
Hence, we compare to the benchmark of \cite{ahmadi2006multi}. Besides, \cite{ahmadi2006multi} uses area partitioning algorithm that relies on
communications and negotiations among robots. Thus, it is prone to message loss and inconsistency. In contrast, our approach
does not require communications among agents, and the partitioning is conducted separately. Hence, we present
a hybrid of centralized and distributed approaches, and the efficacy is demonstrated in the evaluation section.

\section{Problem And Approach}
In this section, we formally define the multi-agent continuous transportation problem in detail, and give
an overview of our approach. In a continuous transportation task, a multi-agent team visits a number of fixed locations and transport objects
to a transportation hub.

\subsection{Formal Problem Definition}
The multi-agent continuous transportation problem can be defined as follows:
\begin{itemize}
\itemsep0em
\item $\mathcal{A}=\{a_1,...,a_{k}\}$ is the set of transportation agents, e.g., the autonomous vehicles.
\item $s_i$, $c_i$, and $y_i\ (\leq c_i)$ denote agent $a_i$'s speed, maximum capacity, and current load, i.e., the number of objects carried.
\item $\mathcal{L}=\{l_1,...,l_{n}\}$ is the set of locations, where $l_0$ is the transportation hub.
\item $v_{j,t}$ denotes the number of objects available at location $l_j$ at time step $t$, e.g.,
the number of passengers appear at the transit stop $l_j$.
\item $\hat{v}_{j,t}^{(i)}$ denotes $a_i$'s estimate of $v_{j,t}$ at time step $t$.
\item $o_j$ denotes the observation made at location $l_j$. When a transportation agent $a_i$ arrives at a location $l_j\ (j > 0)$,
min($v_{j,t}$, $c_i-y_i$) objects at $l_j$ are picked up by $a_i$,
and $a_i$ makes an observation $o_j$ of the number of objects
remaining. When $a_i$ arrives at $l_0$, all $y_i$ objects carried by $a_i$ are transferred to $l_0$.
\item $D:\mathcal{L}\times\mathcal{L}\rightarrow\mathbb{R}^+$ is the distance function of the locations.
\item $t(a_i,l_j,l_k)=\ceil*{\frac{D(l_j,l_k)}{s_i}}$ is the time steps taken for an agent $a_i$ to move from location $l_j$ to $l_k$. 
\end{itemize}

The goal is to maximize the rate of objects delivered to the transportation hub $l_0$ within
$T$ time steps, i.e., maximize $\frac{v_{0,T}}{T}$.

\subsection{Our Approach}
Our approach for solving the continuous transportation problem is:
\begin{itemize}
\itemsep0em
\item We assume that $v_{j,t}$ follows a known model --- in
this paper, we assume that $v_{j,t}$ follows the Poisson model, where the number of objects replenished every time step follows a Poisson distribution with mean $\lambda_j$. However, the parameters of the models (i.e., $\lambda_j$) are not known in advance;
\item The estimates $\hat{v}^{(i)}_{j,t}$ are updated using the model and
observations from transportation agents $a_i$ visiting location $l_j$;
\item Following \cite{liemhetcharat2015continuous}, the transportation agents do not share their models $\hat{v}^{(i)}_{j,t}$ since it could
be expensive subject to the communication bandwidth and team size.
Different from \cite{liemhetcharat2015continuous}, in our approach, the sharing of destination and load among agents are not required.
\end{itemize}
We contribute an online algorithm that partition the
locations based on their 2D-position and estimated replenishment rate, and controls the 
transportation agents $a_i$, that use $\hat{v}^{(i)}_{j,t}$ to plan
their next destination within the cluster of locations assigned. The algorithm dynamically repartitions locations based on information gathered
by the multi-agent team.

\section{Algorithm and Evaluation}
Our Online Balanced Partitioning (OBP) algorithm begins with statically
partition the locations into clusters based on 2D-position of locations with $k$-means algorithm.
The algorithm is inspired by algorithms proposed for continuous foraging
\cite{liemhetcharat2015continuous}. The main difference is
that the agents replan destinations in the cluster of locations assigned, instead of all the locations.
The replanning of destination within each cluster is based on Greedy Rate \cite{liemhetcharat2015continuous}, i.e., maximizing the transportation rate.
Further, due to partitioning, communications of destinations and loads are not required.

By only considering the 2D-position of locations, the workload of each agent might not be balanced, i.e., the total replenishment rate of some clusters could be so high that the agents cannot afford it, while some other agents are idle. On the other hand, the result of standard $k$-means algorithm depends on the choice of initial
centroid which is randomly generated. In this case, we introduce online balanced partitioning, i.e., balance estimated total
replenishment rate of each cluster with information gathering. The replenishment rate is not known in advance. The agents use
preset estimated replenishment rate to update its estimated number of objects. The estimate can only be
corrected when they visit the location and make an observation. Since we focus on
Poisson replenishment model, where number of objects replenished per time step follows a mean value,
we believe that estimated replenishment rate can be corrected with continuous observations,
i.e., the total number of objects replenished divided by time steps elapsed.

Our online partitioning algorithm dynamically corrects the estimated replenishment rate
by visiting a location and making observation. Once the deviation of total
replenishment rate of clusters are greater than a preset threshold, repartitioning will be done immediately.

\begin{figure}[!ht]
  \centering
    \includegraphics[width=0.5\textwidth]{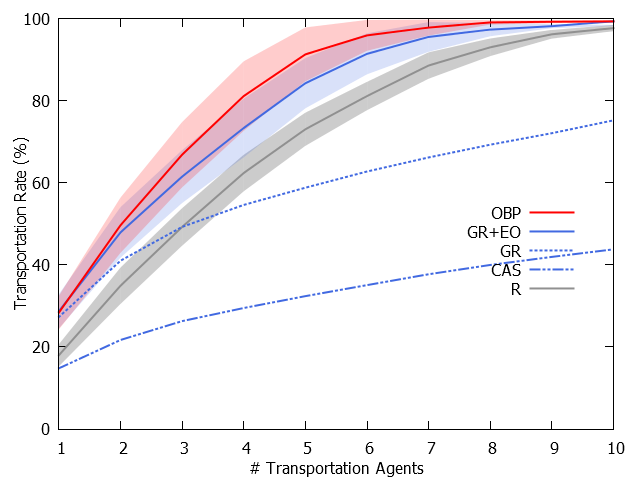}
    \caption{Comparison of our Online Balanced Partitioning (OBP) algorithm against the benchmark of Greedy Rate with Expected Observation (GR+EO) and Continuous Area Sweeping (CAS) algorithm.}
\label{fig:OBP}
\end{figure}

Figure \ref{fig:OBP} shows the performance of our Online Balanced Partitioning (OBP) algorithm when the capacities of
the agents are 10 and the number of locations are 20. In order to demonstrate the effectiveness of OBP, we compared OBP with GR+EO \cite{liemhetcharat2015continuous} which requires
the presence of a reconnaissance agent. The solid red, blue, and gray lines show Online Balanced Partitioning (OBP),
Greedy Rate with Expected Observation (GR+EO), and Random Transportation (R), and the shaded areas show the standard
deviations of these algorithms.

As the number of agents increase, OBP outperforms all other algorithms, even GR+EO ($p=3\times 10^{-23}$),
primarily because of the dynamic partitioning with information gathering by the multi-agent team.
It clearly illustrates the efficacy of our OBP algorithms over GR. Without the use of reconnaissance agent and communications among agents,
our OBP algorithm can still outperform GR+EO algorithm.
\section*{Acknowledgments}
This work was supported by NUS Computer Center
through the use of its high performance computing facilities.
%
%
\bibliographystyle{abbrv}
\bibliography{sigproc}  
%
\end{document}